\documentclass[%
 reprint,
 amsmath,amssymb,
 aps,
]{revtex4-2}

\usepackage{graphicx}
\usepackage{dcolumn}
\usepackage{bm}
\usepackage{caption}
\usepackage{subcaption}
\usepackage{multirow}

\usepackage{hyperref}
\usepackage{url}

\hypersetup{
    colorlinks=true,
    linkcolor=blue,
    filecolor=magenta,      
    urlcolor=cyan,
    }

\begin{document}

\preprint{APS/123-QED}

\title{Simulation of multi-shell fullerenes using Machine-Learning Gaussian Approximation Potential}

\author{C. Ugwumadu}
\email{cu884120@ohio.edu}
\affiliation{Department of Physics and Astronomy, \\
Nanoscale and Quantum Phenomena Institute (NQPI),\\
Ohio University, Athens, Ohio 45701, USA}%

\author{K. Nepal}
\affiliation{Department of Physics and Astronomy, \\
Nanoscale and Quantum Phenomena Institute (NQPI),\\
Ohio University, Athens, Ohio 45701, USA}%

\author{R. Thapa}
\affiliation{Department of Physics and Astronomy, \\
Nanoscale and Quantum Phenomena Institute (NQPI),\\
Ohio University, Athens, Ohio 45701, USA}%

\author{Y. G. Lee}%
\affiliation{Department of Physics and Astronomy, \\
Nanoscale and Quantum Phenomena Institute (NQPI),\\
Ohio University, Athens, Ohio 45701, USA}%

\author{Y. Al Majali}%
\affiliation{Russ College of Engineering and Technology, \\
Ohio University, Athens, Ohio 45701, USA}%

\author{J. Trembly}%
\affiliation{Russ College of Engineering and Technology, \\
Ohio University, Athens, Ohio 45701, USA}%

\author{D. A. Drabold}%
\email{drabold@ohio.edu}
\affiliation{Department of Physics and Astronomy, \\
Ohio University, Athens, Ohio 45701, USA}%

\date{\today}

\begin{abstract}
Multi-shell fullerenes "buckyonions" were simulated, starting from initially random configurations, using a density-functional-theory (DFT)-trained machine-learning carbon potential within the Gaussian Approximation Potential (GAP) Framework [Volker L. Deringer and Gábor Csányi, Phys. Rev. B 95, 094203 (2017)]. Fullerenes formed from seven different system sizes, ranging from 60 $\sim$ 3774 atoms, were considered. The buckyonions are formed by clustering and layering starting from the outermost shell and proceeding inward. Inter-shell cohesion is partly due to interaction between delocalized \text{$\pi$} electrons protruding into the gallery. The energies of the models were validated \textit{ex post facto} using density functional codes, \texttt{VASP} and \texttt{SIESTA}, revealing an energy difference within the range of 0.02 - 0.08 eV/atom after conjugate gradient energy convergence of the models was achieved with both methods.

\end{abstract}

\keywords{Carbon; Buckyonion; Fullerenes; Machine learning; Gaussian Approximation Potential}

\maketitle


\section{\label{sec:introduction}INTRODUCTION}

 Research on various allotropes of Carbon (C) like diamond, graphite, nanotubes, and fullerenes has been a long-standing endeavor. In recent years, with research shifting to disordered C structures, new carbon materials are being unveiled. Examples of such materials include tetrahedral amorphous Carbon \cite{taC1,taC2}, Biphenylene carbon sheets \cite{Biphenylene_network}, and amorphous graphite (layers of amorphous graphene sheets with topological defects in the planes) \cite{LAG}. It is known that topological defects present unique and interesting mechanical, structural, thermal, and electronic properties that often have scientific and technological consequences. For instance, Zhang and co-workers discovered amorphous carbon structures with a high fraction of sp$^3$ bonding, recovered from compression of solid C$_{60}$ under high pressure and high-temperature \cite{strongest_aC}. The resulting material showed remarkable mechanical and electronic properties, with potential application in photovoltaic systems, requiring high strength and wear resistance. The growing interest in amorphous carbon structure raises the question: what other amorphous carbon materials exist?

\begin{figure*}[!htpb]
	\centering
	\includegraphics[width=.8\linewidth]{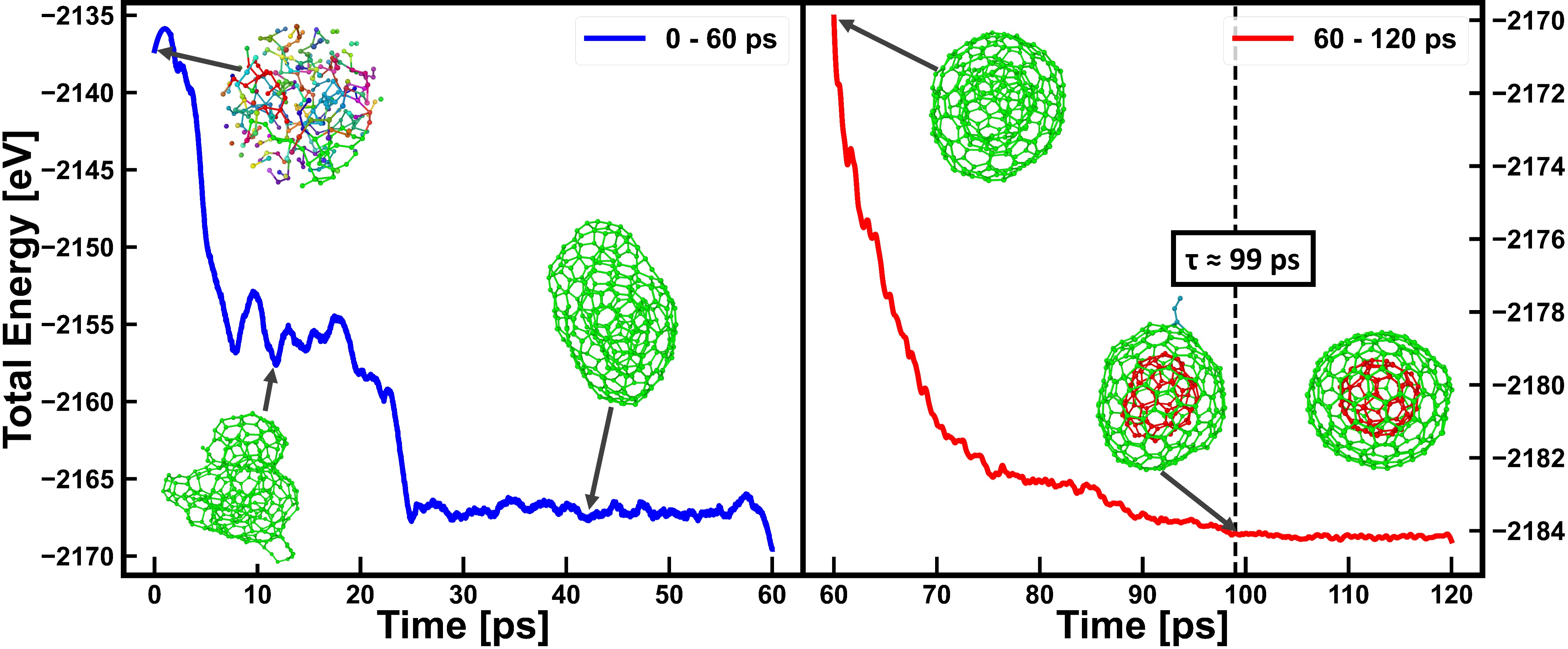}
	\caption{ Shell formation and total energy  as a function of simulation time. The insets (shells are colored green and red) show the snapshots of the atomic configurations at different points in time. $\tau$ indicates the time at which the shells become visible with significant energy convergence. The right plot (60 - 120 ps) is simply a continuation of the left plot (0 - 60 ps).}
	\label{fig:Cfig_MD_details.jpg}
\end{figure*}

 Topologically, there are two kinds of fullerenes, classical and non-classical fullerenes. The classical fullerenes obey the isolated pentagon rule \cite{kroto1987stability} while their non-classical counterparts do not, and in addition to the hexagon rings, non-classical fullerenes contain 4 - 8 member rings. Classical fullerenes, especially C$_{60}$, have been applied in fuel cell catalysis \cite{fuelcell}, production of supercapacitors and lithium-ion batteries \cite{supercapacitor, lithiumcell},  electromagnetic inference shielding\cite{em}, biomedical sciences \cite{medicine}, and even more recently in nano-neuroscience \cite{neuroscience}. Applications using non-classical fullerenes are limited by their unique physical and chemical properties, which is a consequence of the variation in their local atomic conformations resulting from the inclusion of the non-hexagonal rings. It becomes apparent that an atomistic understanding of these amorphous fullerenes could present an opportunity for targeted engineering of their structures for advanced technological applications. Research on the synthesis, as well as theoretical study on non-classical single-shell fullerenes, has been reported in a number of publications \cite{nc1,nc2,nc3,nc4}. On the other hand, limited theoretical research exists for non-classical multi-shell fullerenes (``buckyonions"). Any progress in this area would require a fundamental understanding of the formation and inter-layer cohesion mechanism in non-classical buckyonions through molecular dynamics (MD) simulations that allow sampling from an extensive statistical ensemble of the configuration space that are energetically favorable. Some notable research on MD simulation of non-classical buckyonions (but using ``precursors") include the work of Los and co-workers \cite{ancestor1}, who simulated the mechanism of buckyonion formation from the annealing of ultra-disperse nano-diamonds at 3000 K using the classical long-range carbon bond-order potential (LCBOPII) \cite{LCBOPII}. Also, Lau and co-workers \cite{ancestor2} investigated the micro-structure of buckyonions formed using pulsed-laser deposition by annealing an a-C structure at 4000 K, using the environment-dependent interaction potential (EDIP) for carbon \cite{EDIP}. 

 This work reports the formation of buckyonions without any precursor. Starting from a random configuration of carbon atoms with different system size ranging from 60 $\sim$ 3774 atoms, non-classical multi-shell fullerenes (with up to 4 shells) were formed using a density-functional-theory trained machine-learning Gaussian Approximation Potential (GAP) within 40 $\sim$ 130 ps. The buckyonion models formed were validated through DFT energy minimization in \texttt{VASP} (Vienna Ab initio Simulation Package) \cite{VASP} and the atomic orbital-based package, \texttt{SIESTA} (Spanish Initiative for Electronic Simulations with Thousands of Atoms) \cite{SIESTA}. The multi-shell fullerenes formed up to four shells with topological disorder in each shell (that is, 3D aspherical structures with hexagons, pentagons, heptagons, etc.). Besides a thorough structural investigation with a highly accurate potential, we study mechanisms of formation and the atomistic origins of cohesion in detail. For where it applies, two exchange-correlation functionals were used for the DFT calculations within \texttt{VASP}. These are the generalized gradient approximation (GGA) of Perdew-Burke-Ernzerhof (referred to as PBE) \cite{PBE} and the combination of the slater exchange \cite{Slater} with the Perdew-Zunger parametrization of Ceperley-Alder Monte-Carlo correlation data in the local density approximation \cite{LDA1,LDA2} (referred to as LDA). However, except where explicitly stated, the results discussed herein are those obtained using the PBE functional.

\section{\label{sec:Computational_details}Computational Details}

  The models were obtained using the GAP potential for the C-C interaction \cite{C} as implemented within the ``Large-scale Atomic/Molecular Massively Parallel Simulator" (LAMMPS) software package \cite{lammps}. The training of the GAP C potential was based on the conventional method of fitting to an extended set of reference data obtained from simulations based on density functional theory (DFT) \cite{gap2,C2,C3,C4}. Three initial configurations with 60, 300, 540, 840, 1374, 2160, and 3774 atoms were randomly placed to form a spherical bulk with all C atoms being at least 1.45 \AA~apart. These systems will henceforth be referred to as BO$_N$, where \textit{N} is the number of atoms. The bulk was then placed into a cubic box with 3 \AA~vacuum between the surface of the sphere and the sides of the box so that the periodic boundary condition describes a system of isolated carbon clusters. The atomic positions and velocities were sampled and updated in time-steps of 1 fs from a Nos\'e–Hoover thermostat at a fixed temperature of 3000 K. The energy trajectories of the systems were analyzed and taken to have reached a satisfactory configuration after negligible energy fluctuations were observed over an extended time period of at least 10 $\sim$ 15 ps. Afterwards, the models were allowed to find a more energetically and structurally favorable configuration by cooling to room temperature at a rate of 2.7$\times$10$^{13}$ K/s, this allowed the systems to attain nearly spherical geometry. Finally, the structures were relaxed using conjugate gradient (CG) algorithm as implemented within LAMMPS with a force tolerance of 10$^{-6}$ eV/\AA\ to obtain a representative structure. Figure \ref{fig:Cfig_MD_details.jpg} presents the time-evolution of the BO$_{300}$ model from one of the initial random configurations to an energetically converged two-layer buckyonion structure. The parameter $\tau$ indicates the time at which the concentric shells become discernable with essentially non-fluctuating energies. While most of this work was carried out with the aforementioned systems, models with other $N$ were explored for some of the analysis. 

 The structure and energy of the BO$_N$ models obtained were validated \textit{ex post facto} using \texttt{VASP} and \texttt{SIESTA}. To achieve this, the systems were energetically relaxed using CG algorithm in \texttt{SIESTA} and \texttt{VASP}. Fig. S1 in the supplementary material \cite{suppl} shows the plots of the energy difference ($\Delta E$) against the CG iterations for BO$_{300,540,840}$, obtained from the density functional codes. The inset in Fig. S1, representing the GAP model before and after CG relaxation, showed no structural difference in the relaxed models when compared to the original models. The energy difference between the original models and relaxed models (both \texttt{SIESTA} and \texttt{VASP}) ranged within $0.02~\text{eV/atom} \leq \Delta E \leq 0.08~\text{eV/atom}$ also with no bond breaking or forming. This is an indication that in the configuration space, the atoms are formed in an energetically stable and realistic local environment.  The “machine learning” process involves the absence of an ansatz and a fit of the training (DFT) data to a flexible function. This allows GAP to scan through a wider range of configuration space for a realistic description of structurally complex systems like the buckyonions. In principle, one could model buckyonions from random configurations using \textit{ab-initio} DFT codes like \texttt{VASP}, but this becomes impractical as a result of the plane wave basis set which scales with the cell volume (that contains vacuum). GAP is a real-space method, which is an advantage for molecular systems and also has a computational cost that scales linearly with the system size.
 
\section{\label{sec:structure} Structural Analysis} 

\begin{table*}[!t]
	\begin{center}
		\caption{Simulation parameters for the buckyonion models.}\label{tab:Ctable1}
		\begin{tabular*}{0.9\linewidth}{@{\extracolsep\fill}cc c|cccc|}
			\hline
			 &  & &
			\multicolumn{4}{c|}{\makebox[0pt]{\textbf{layer Size [number of atoms]}}} \\
			\textbf{Model} & \textbf{$\tau$ [ps]} &  \textbf{Layers} & $s_1$  & $s_2$ & $s_3$ & $s_4$ 
			\\\hline

 			BO$_{60}$       & 41     & one    & 60     & -    & -  & -  \\
			BO$_{300}$      & 99     & two    & 76     & 224 & - & -   \\
			BO$_{540}$      & 104     & two   & 174     & 366 & - & -   \\
			BO$_{840}$      & 119    & three &  25    & 282    & 533  & -  \\
			BO$_{1374}$     & 109     & three & 158   & 456    & 760  & -    \\
			BO$_{2160}$     & 113    & three  & 189     & 799    & 1172  & -     \\
			BO$_{3774}$     & 131    & four  & 316     & 572    & 1210    & 1676   \\
			\hline
		\end{tabular*}
	\end{center}
\end{table*}
\begin{table*}[t!]
	\begin{center}
		\caption{Shape descriptors for the buckyonion models.}\label{tab:Ctable2}
		\begin{tabular*}{0.9\linewidth}{@{\extracolsep\fill}c|cccc|cccc|ccc|}
			\hline
			 & 
			\multicolumn{4}{c|}{\makebox[0pt]{\textbf{Gyration radius (R$_g$) [\AA]}}}  &
			\multicolumn{4}{c|}{\makebox[0pt]{\textbf{Asphericity ($\eta$) [\AA$^2$]}}}    
			& 
			\multicolumn{3}{c|}{\makebox[0pt]{\textbf{Anisotropy ($\kappa^2$)}}} \\
			\cline{2-12}
			\textbf{Model}   & $s_1$  & $s_2$ & $s_3$ & & $s_1$  & $s_2$ & $s_3$ & & $s_1$  & $s_2$ & $s_3$
			\\\hline

			BO$_{60}$       & 3.57  & -     & -     & & 0.525    & -     & -     & & 0.18     & -    & -        \\
			BO$_{300}$      & 4.05  & 6.89  & -     & & 0.27    & 0.43  & -     & & 0.03      & 0.01  & -      \\
			BO$_{540}$      & 6.06  & 8.9  & -     & &0.71   & 0.74  & -     & & 0.14     & 0.07  & -      \\
			BO$_{840}$      & 2.4     & 7.76     & 10.63     & & 1.48       & 1.16  &  1.24   & & 0.12         &1.89     & 0.11       \\
			BO$_{1374}$     & 5.84     & 9.74     & 12.55     & & 0.85       & 0.87     & 0.92   & & 0.19         & 0.03     & 0.02      \\
			BO$_{2160}$     & 5.88     & 12.79     & 15.60     & & 1.69      & 1.41   & 1.43   & & 0.18         & 0.06     & 0.04      \\
			BO$_{3774}$     & 8.19     & 11.03     & 16.01     & & 0.46       & 0.48     & 0.85   & & 0.01         & 0.003     & 0.005      \\\hline
			C$_{60}$        & 3.44     & -     & -     & & 0.01         & -     & -     & & $\approx$ 0         & -     & -      \\
			SWCNT        & 12.78    & -    & -    & & 144.7      & -     & -     & & 0.99        & -    & -       \\\hline
			BO$_{3774}$     &      & $s_4 = 18.90$    &     & &        &$s_4 = 0.87$     &     & &          & $s_4 = 0.004$    &       \\\hline
		\end{tabular*}
	\end{center}
\end{table*}

\begin{figure}[!t]
	\centering
	\includegraphics[width=.8\linewidth]{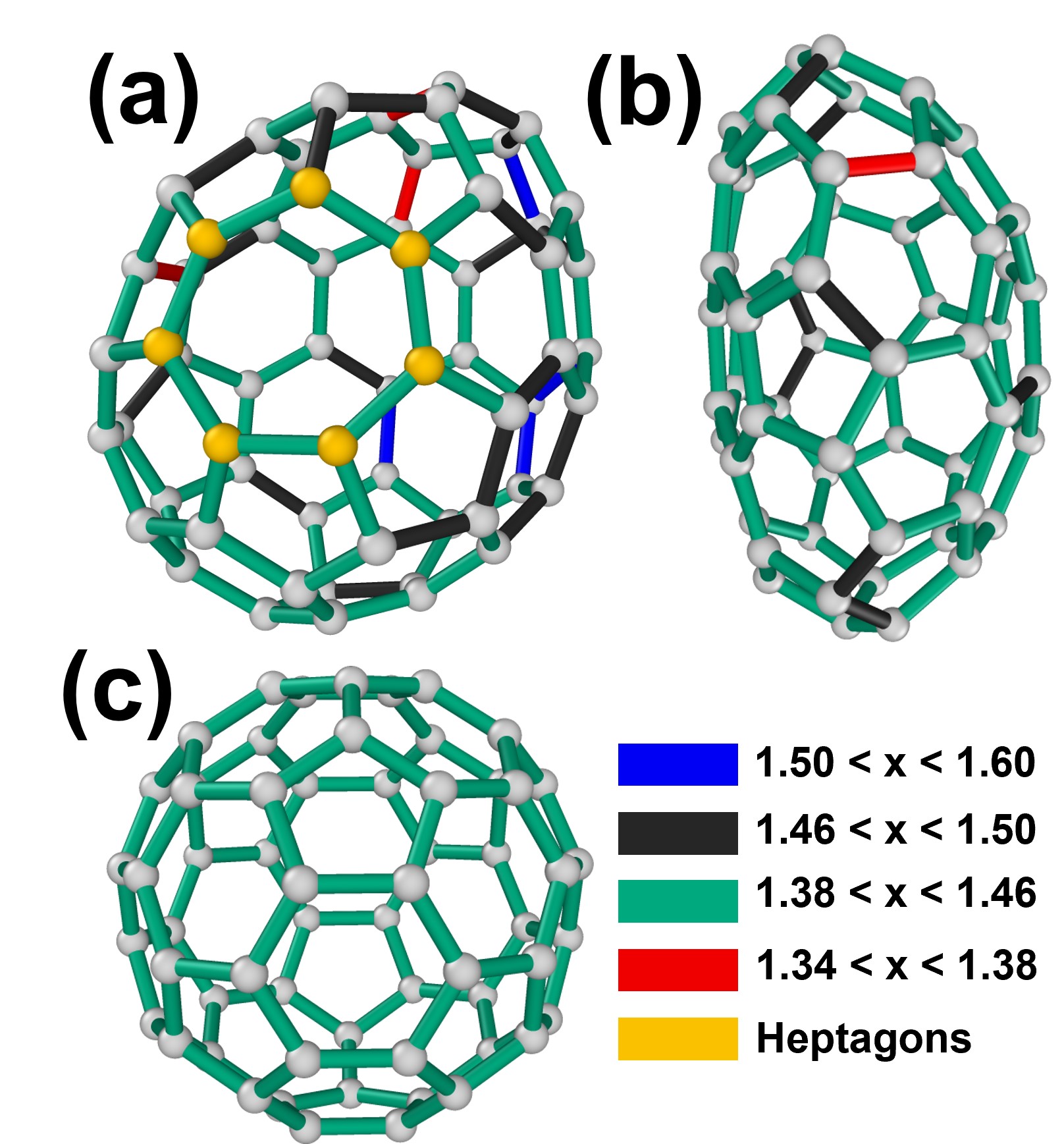}
	\caption{Topological defects observed in BO$_{60}$ models (a and b). x represents the calculated bond-lengths in \AA. The models are compared with a pristine C$_{60}$ (C) with I$_h$ symmetry \cite{K_F60}}
	\label{fig:Cfig2_Structures.jpg}
\end{figure}

Tables \ref{tab:Ctable1} and \ref{tab:Ctable2} contain information  about the simulation parameters and some structural properties calculated for the buckyonion models respectively. Generally, two families of icosahedral multi-shell fullerenes exist, the  60$n^2$ and 180$n^2$ buckyonions; hence, the smallest shell-unit for buckyonions would be the Buckminsterfullerene (C$_{60}$) \cite{typesofBO}. Variations in the C-C bond-length, as well as Stone-Wales defects in fullerenes, could cause the re-ordering of atoms in the buckyonion layers and yield deviations from the expected shape of the fullerenes. This topological disorder is present in all the buckyonion models and indicated in Fig. \ref{fig:Cfig2_Structures.jpg} (a) and (b) for two independent BO$_{60}$ models. Unlike in C$_{60}$ (see Fig. \ref{fig:Cfig2_Structures.jpg}c), which has the expected C-C bond-length ($1.38~\text{\AA} < x < 1.46~\text{\AA} $), the BO$_{60}$ and other buckyonions have C-C bond-length ranging from 1.34 \text{\AA} $\sim$ 1.6 $\text{\AA}$. Furthermore, energy calculation using GAP revealed that the BO$_{60}$ model was 0.03 eV/atom higher in energy compared to pristine C$_{60}$. Similar calculations for other models are presented in Table \ref{tab:Ctable_EAtom}.

\begin{figure*}[!htb]
   \centering
        \includegraphics[width=\linewidth]{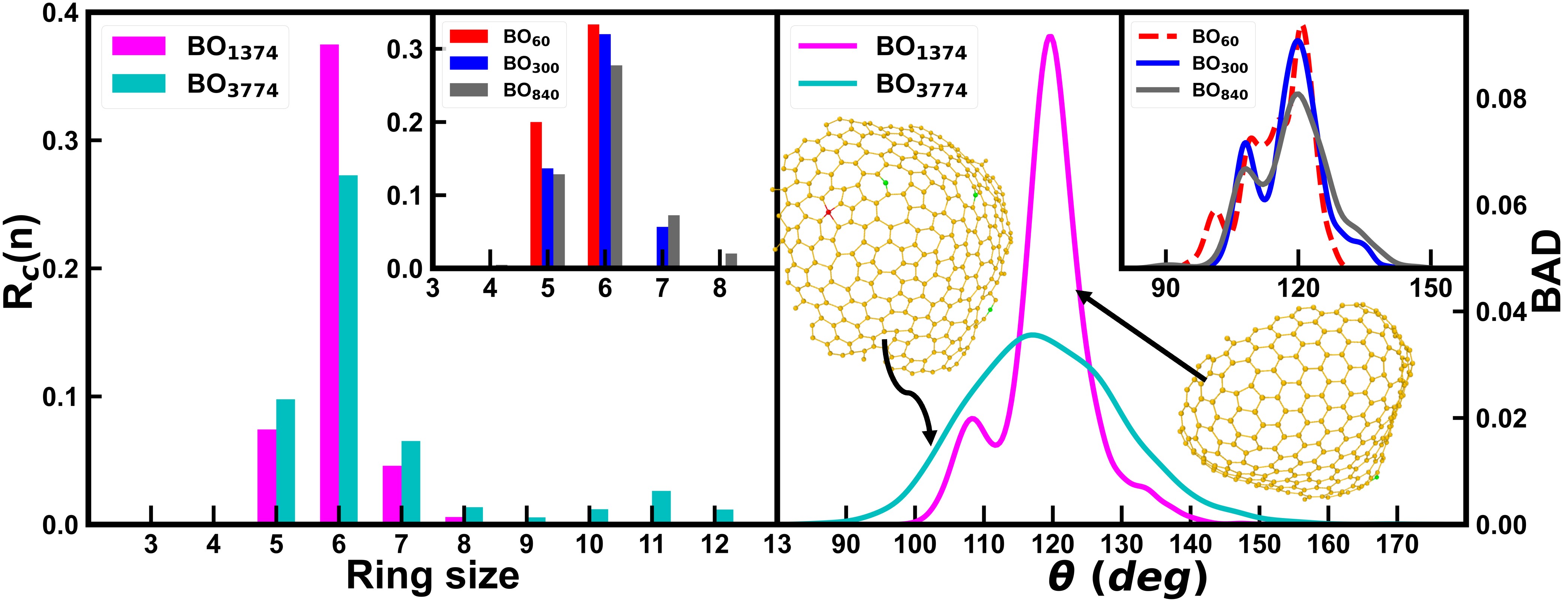}
        \caption{Ring statistics (LEFT) and bond-angle distribution (RIGHT) for the local environment of the buckyonion models as calculated within ISAACS \cite{isaacs}. The honeycomb structure in the inset is a representative structure for some atoms in BO$_{1374}$ and BO$_{3774}$ models.}
        \label{fig:Cfig_ring_bad}
\end{figure*}

The histograms in Fig. \ref{fig:Cfig_ring_bad} (LEFT) represent the ring distribution for some of the BO$_N$ models with the corresponding bond-angle distribution shown in Fig. \ref{fig:Cfig_ring_bad} (RIGHT). For BO$_{1374}$, the peaks at 109$^{\circ}$  and 119$^{\circ}$ relate directly to the pentagonal and hexagonal atomic arrangements (inset shows a portion of BO$_{1374}$ representative shell with 5- and 6-member rings). On the other hand, the broadening of the angle distribution in  BO$_{3774}$ shows variations in ring size. A comprehensive description of the number of ring sizes found in the models is presented in Table S1 \cite{suppl}.

    \begin{table}
	\caption{Difference in Energy per atom  between the buckyonion models and C$_{60}$  calculated using GAP ($\delta E_\text{atom} = E_{BO_{N}} - E_{C_{60}}$).}
		\label{tab:Ctable_EAtom}
	\begin{ruledtabular}
		\begin{tabular*}{\linewidth}{@{\extracolsep\fill}cc|cc@{\extracolsep\fill}}
			\textbf{Models}&\textbf{$\delta E_\text{atom}$ [eV]}&
			\textbf{Models}&\textbf{$\delta E_\text{atom}$ [eV]}\\
			\hline
			 BO$_{60}$& \ 0.03   &  &    \\
			 BO$_{300}$& - 0.30 & BO$_{1374}$& - 0.41     \\
    		 BO$_{540}$& - 0.30  &  BO$_{2160}$& - 0.43  \\
			 BO$_{840}$& - 0.31  & BO$_{3774}$& -   0.38   \\
		\end{tabular*}
	\end{ruledtabular}
\end{table}
    

\begin{figure}[!ht]
	\includegraphics[width=\linewidth]{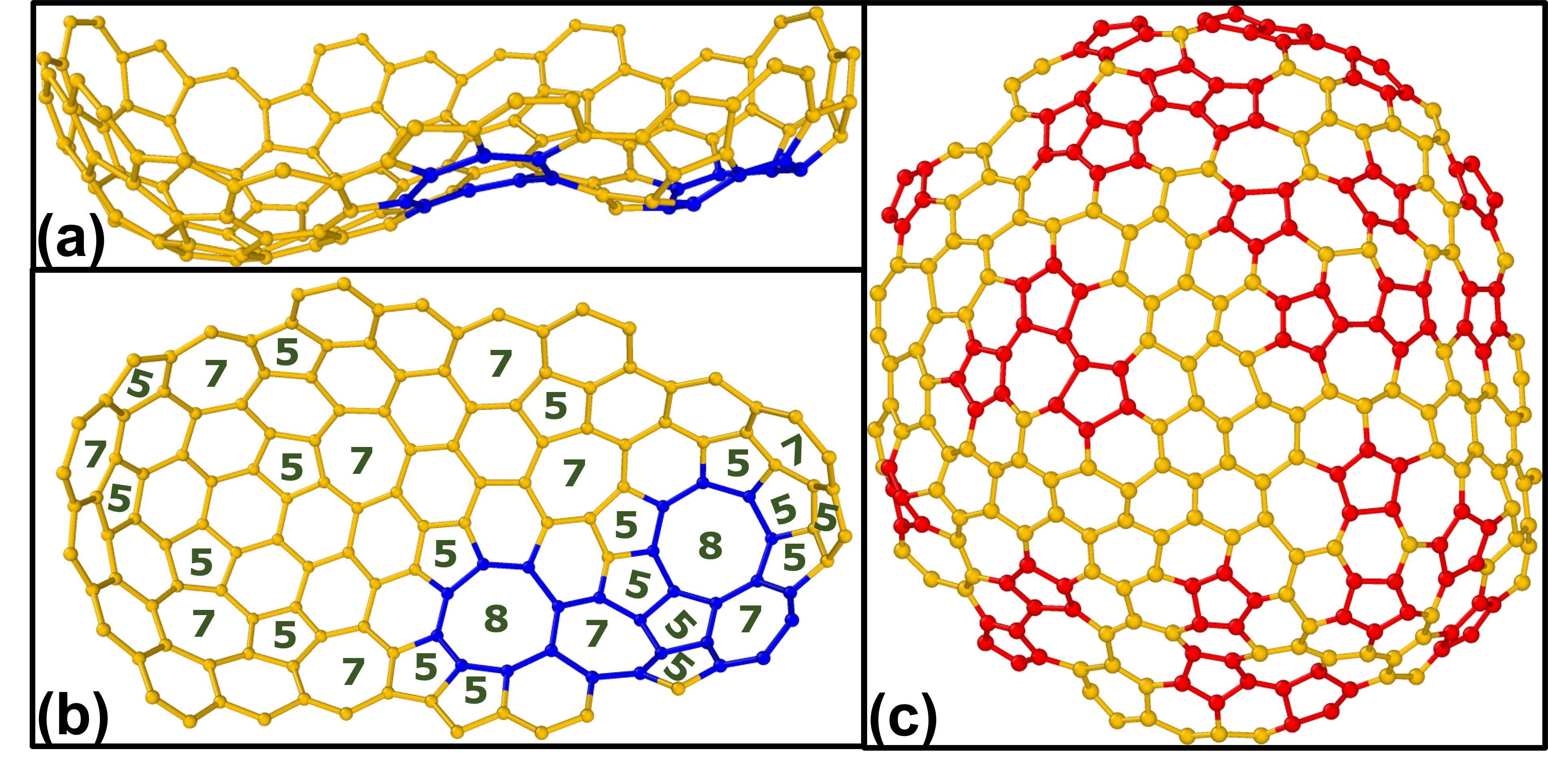}
	\caption{The surface curvature induced by ring sizes. (a) shows the negative curvature defined by the presence of octagons (blue) and heptagons in (b). (c) shows a positive curvature induced by a large concentration of pentagon (red) in some representative surfaces of the buckyonion model.}
	\label{fig:CSfig_1347_curve}
\end{figure}

A pentagonal (heptagonal and octagonal) ring in graphitic carbon sheet results in a positive (negative) curvature \cite{Mackay,TERRONES,Lenosky,vanderbuilt,schwarziteDAD,schwarziteHomyonfer,TAGAMI}. However, combinations of  5-, 7- and/or 8-member rings in hexagonal carbon networks could result in a complicated morphology \cite{IIJIMA,Wang_exp}. Fig. \ref{fig:CSfig_1347_curve} (a) and (b) show a part of the BO$_{840}$ model with a negative curvature (the 7/8 pairs are colored in blue), while Fig. \ref{fig:CSfig_1347_curve} (c) shows a positive curvature region in BO$_{1374}$ with a high number of pentagons (in red). All the BO$_N$ structures are displayed in Fig. S2 \cite{suppl}. To form fullerenes, Euler's polyhedral formula requires a balance in the number of 5/p pairs (p represents 7- or 8- member ring) for an indeterminate number of hexagon rings \cite{Lenosky, Naito}, and any deviation from the 5/p pair at a certain region must be compensated by additional 5-member rings if a positive curvature is to be maintained as shown in Fig. \ref{fig:CSfig_1347_curve} (b). 


Even with defects, the number of atoms in the innermost shell for the 300-atom and 520-atom models remained close to 60 and 180 atoms respectively (see Table \ref{tab:Ctable1}), and maintained a nearly spherical geometry in comparison to large $N$ models which are more faceted. This supports experimental evidence that small $N$ buckyonions exist as concentric spherical shells and as faceted structures as $N$ increases \cite{T12,T13,T14,E1,E2,E3}.

Further analysis on the shape of the BO$_N$ models was carried out using various shape descriptors derived from the gyration tensor (\textbf{S}$_{ij}$; $i,j = x,y, z$) for each buckyonion shell. The squared radius of gyration (R$^2_g$) was calculated as the sum of the principal moments ($\lambda_i^2$; $i = x,y,z$) of the gyration tensor as:
 
  \begin{equation}
     \label{eq0}
     R^2_g = \sum_{i=x,y,z} \lambda_i^2 
 \end{equation}
 
 \noindent where the axes were chosen in a way that the principal moments are ordered as $\lambda_z \ge \lambda_y \ge \lambda_x$
 
 The spherical symmetry of each shell in the buckyonion models was calculated using the asphericity descriptor ($\eta$) given as:
 
 \begin{equation}
     \label{eq1}
     \eta = \frac{3}{2} \lambda_z^2 - \frac{R_g^2}{2}
 \end{equation}
 
 \noindent which is non-negative and is only zero (spherically symmetric) when $\lambda_z = \lambda_y = \lambda_x$. The  relative shape anisotropy ($ \kappa^2 \in (0,1)$) was also calculated using the formula:
 
 \begin{equation}
     \label{eq3}
     \kappa^2 = \frac{\eta^2 + (3/4)\zeta^2}{R_g^4}
 \end{equation} 
 
 \noindent where $\zeta =\lambda_y^2 - \lambda_x^2$. $\kappa^2$ is 0 for perfectly spherical systems is and it is 1 for particles (atoms) that lie on a line. Similar calculations were done for the C$_60$ fullerene and a single-walled nanotube to compare with the BO$_N$ model. The results from these calculations are presented in Table \ref{tab:Ctable2}. 

\begin{figure*}[!t]
	\centering
	\includegraphics[width=\linewidth]{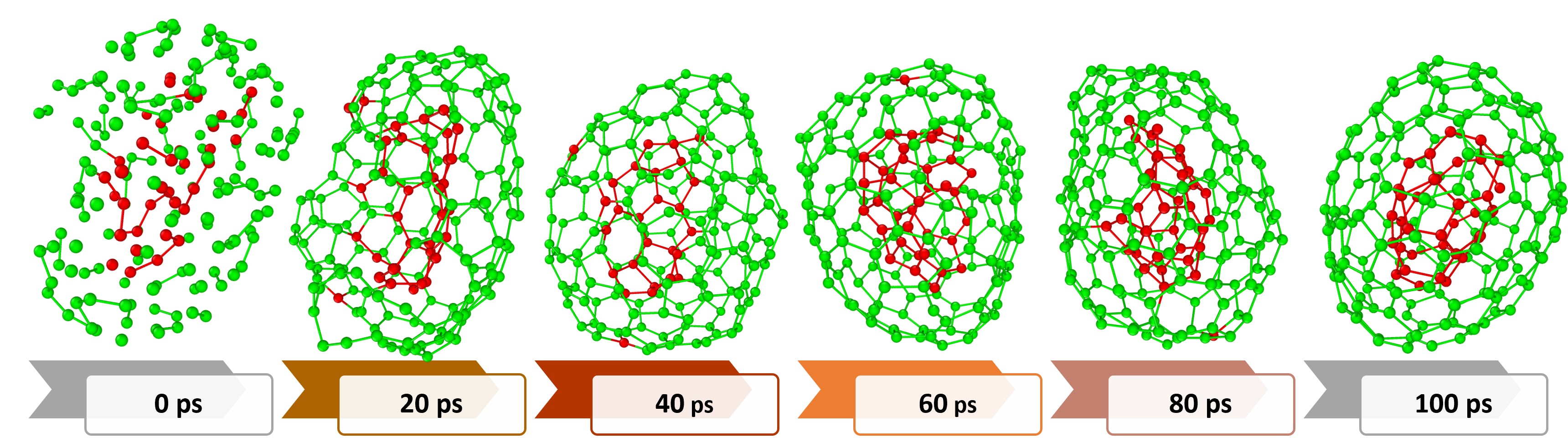}
    	\caption{Clustering Process for buckyonion  formation from two bowl-shaped, randomly distributed clusters with 36- (red) and 152- (green) atoms.}
	\label{fig:Cfig3_clusters.jpg}
\end{figure*}

Small-angle neutron scattering (SANS) is a valuable technique for structural characterization of fullerenes in solvents with strong SANS contrast (e.g., CS$_2$) \cite{sans1,sans3,sans2,sans4,sans5}. The structures of some BO$_N$ ($N$ = 60, 70, and 84) were compared with corresponding structures observed experimentally from the work of Melnichenko and co-workers \cite{Melnichenko}. Using SANS, they analyzed the shapes of fullerenes with $N= 60, 70$ and $84$ moieties by determining the radius of gyration (R$_g$) via the standard Guinier approximation of the low-Q end of the scattering profile given as \cite{guinier}:

 \begin{equation}
     \label{eq4}
     I(Q) \approx I(0)\cdot \exp \left(\frac{-(QR_g)^2}{3}\right)
 \end{equation}

\noindent $I(0)$ is the intensity at zero scattering angle ($Q=0$). R$_g$ and I(0) are determined from the linear fit to a plot of $ln~(I)$ against $Q^2$ within the “Guinier region” \cite{guinier_region}. Table \ref{tab:Ctable_compareRg} presents values of the radius of gyration from experimental data \cite{,Melnichenko,sans2,sans4,sans5}, theoretical calculations in ref \cite{Melnichenko}, and those obtained as part of this work for fullerenes with $N$ = 60, 70 and 84. 

 For the 60 atom systems, the result for R$_g$ obtained from this work is consistent with experimental results from the works of Gripon \textit{et al.}\cite{sans2}, Smorenberg \textit{et al.} \cite{sans4}, Spooner \textit{et al.} \cite{sans5} and those from Melnichenko and co-workers. However, for the  79- and 84- atom systems, experimental data was only available from Melnichenko and co-workers, who also calculated a theoretical radius of gyration from the atomic coordinates of the fullerenes using the Serena Molecular Modeling Software \cite{SERENA}. While there is limited experimental data for structure characterization analysis of buckyonions, the ability to reproduce comparable R$_g$ for available fullerenes still serves as a validation for the structure of the BO$_N$ models.
 
 The process of buckyonion formation by clustering was investigated by creating two randomly distributed hemispherical clusters with 36 atoms in the first cluster (red atoms in Fig. \ref{fig:Cfig3_clusters.jpg}) and 152 atoms in the second cluster (green atoms in Fig. \ref{fig:Cfig3_clusters.jpg}). The systems were separated by a distance of at least 3.0 \AA~and the simulation protocol was the same as described in Sect. \ref{sec:Computational_details}.  Fig. \ref{fig:Cfig3_clusters.jpg} shows that there was hardly any mixing between atoms in both clusters, ie. the inner shell was formed majorly by atoms in the 36-atom system and the outer shell was formed by atoms in the 152-atom system. This suggests that, even with random configurations, if clustering exists within the system, the buckyonion will likely form based on those clusters.

\begin{table}[!ht]
	\begin{center}
		\caption{$R_g$ [\AA] for fullerenes with $N=$ 60, 70, and 84 obtained from experimental and theoretical (calculated from atomic coordinates in ref \cite{Melnichenko}) data, compared with corresponding models created using the GAP Potential}\label{tab:Ctable_compareRg}
		\begin{tabular*}{0.9\linewidth}{@{\extracolsep\fill}cccccc@{\extracolsep\fill}}
			\hline
			 & \textbf{60 atoms} & \textbf{70 atoms} & \textbf{84 atoms}\\
			\hline
			\textbf{Exp.}~~\cite{sans2} & $3.5 \pm 0.20$   & - & - \\
			
			\textbf{Exp.}~~\cite{sans4} & $3.52\pm 0.04$   & - & - \\
			
			\textbf{Exp.}~~\cite{sans5} & $3.57\pm 0.00$   & - & - \\
			
			\textbf{Exp.}~~\cite{Melnichenko} & $3.56\pm 0.05$   & $3.92 \pm 0.05$ & $4.27 \pm 0.06$ \\

			\textbf{Theo.}~\cite{Melnichenko} & 3.55 & 3.87 & 4.19 \\
			 \textbf{This work} & 3.57  & 3.78  & 4.05 \\\hline
		\end{tabular*}
	\end{center}
\end{table}

\begin{figure} [!htb]
    \centering
        \includegraphics[width=\linewidth]{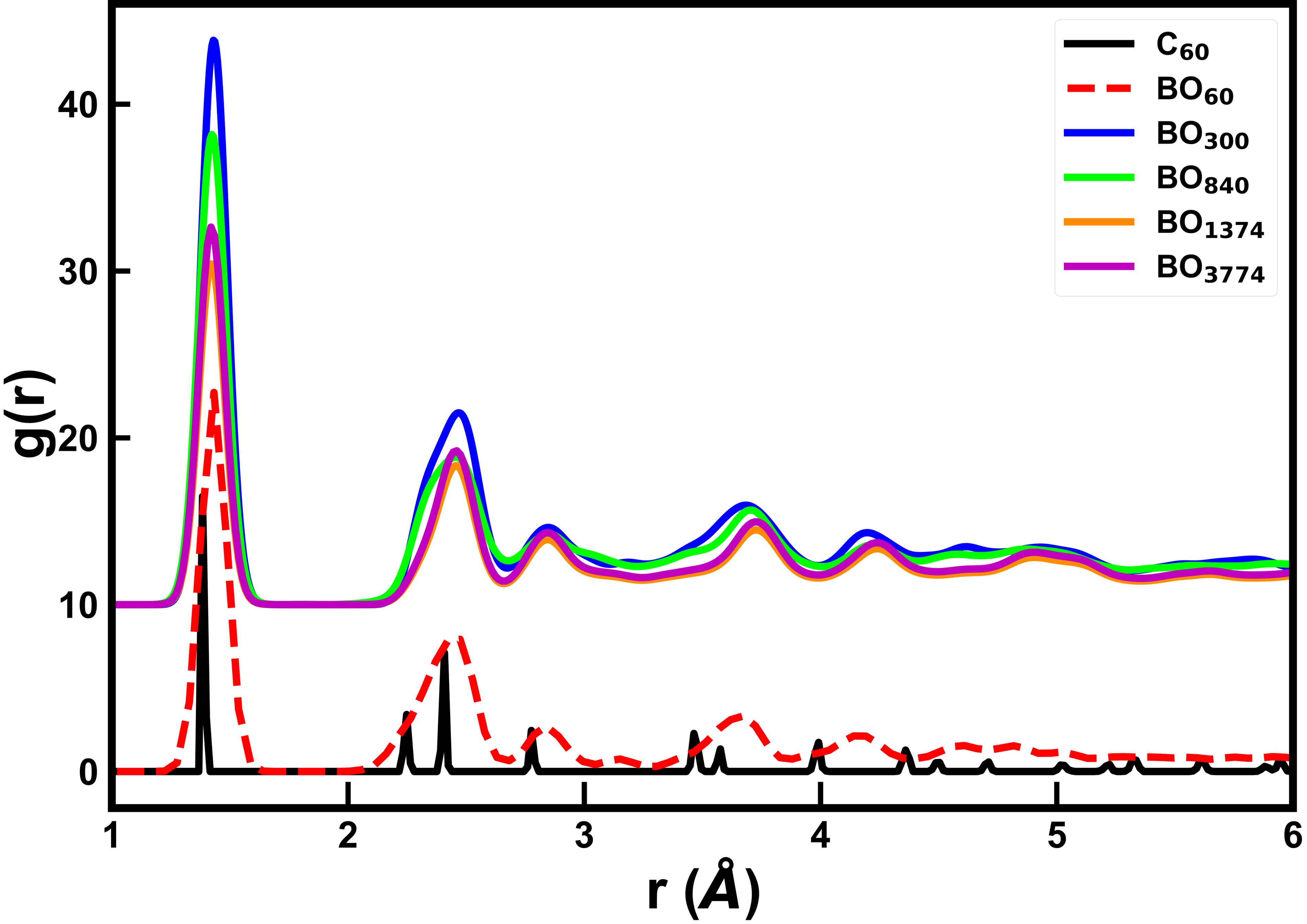}
        \caption{Radial distribution function g(r) for C$_{60}$ fullerene (in black) and some of the BO$_N$ models. Plots were vertically shifted for clarity.}
        \label{fig:Cfig_rdf}
\end{figure}

Next, the structural order of the BO$_N$ models and pristine C$_{60}$ \cite{K_F60} were analyzed using pair-correlation functions. The first peak in the plot for the radial distribution functions for all the configurations in Figure \ref{fig:Cfig_rdf} is centered around the ``graphitic" bond-length and the width of the peak indicates disorder-induced deviations from the bond-length. The continuous peaks in the BO$_N$ models overlap with the discrete peaks in C$_{60}$, which is a validation of the amorphous structure of the buckyonion models. 

The gallery width between any two concentric shells was calculated as the difference of the radii between two concentric shells measured from the centroid of the system. The inter-layer separation between the outermost concentric shell showed a consistent value around  2.87 \AA~for all the models. This value is less than the inter-layer separation observed in pristine Carbon nano-onions ($\approx$ 3.5 \AA) \cite{BO1, BO2, BO3}. In the animation for the formation of BO$_{300}$  and BO$_{1374}$  provided in the supplementary material \cite{suppl}, the path of buckyonion formation brings about the creation of the outermost layer, and then the system slowly forms inner shells. The animation for BO$_{1374}$ confirms that an initial cluster of atoms exists, which is then shaped spherically, followed by the formation of an outermost layer (green shell) with the inner (still random) cluster separated from the outermost shell. Next, a second layer is formed (red shell), which achieves a stable geometry after a considerable amount of time. After this stage has been completed, the formation of the innermost (blue shell) layer commences. The formation process, combined with the consistent gallery width between the two outermost shells, suggests a level of stability in the outer-shell formation process of the buckyonions. The stability of the outer-shell and the fullerene growth process was further confirmed by creating a randomly distributed 540-atom cluster inside a 720-atom spherical fullerene isomer \cite{K_F60}, following the simulation  process in Sect. \ref{sec:Computational_details}. Fig. S3 \cite{suppl} shows the evolution of the system with observable faceting occurring at regions with pentagons in the outermost shell. Analysis of the number of atoms in each shell revealed that the outermost shell remained with 720 atoms (green shell) throughout the simulation while forming two inner-shells with 364 (red shell) and 176 (blue shell) atoms. This is shown in the animation for the growth process (BO growthProcess.mp4) provided in the supplementary material \cite{suppl}. 

\section{Electronic Structure and Charge Density Distribution}

 \begin{figure}[!htpb]
	\centering
	\includegraphics[width=\linewidth]{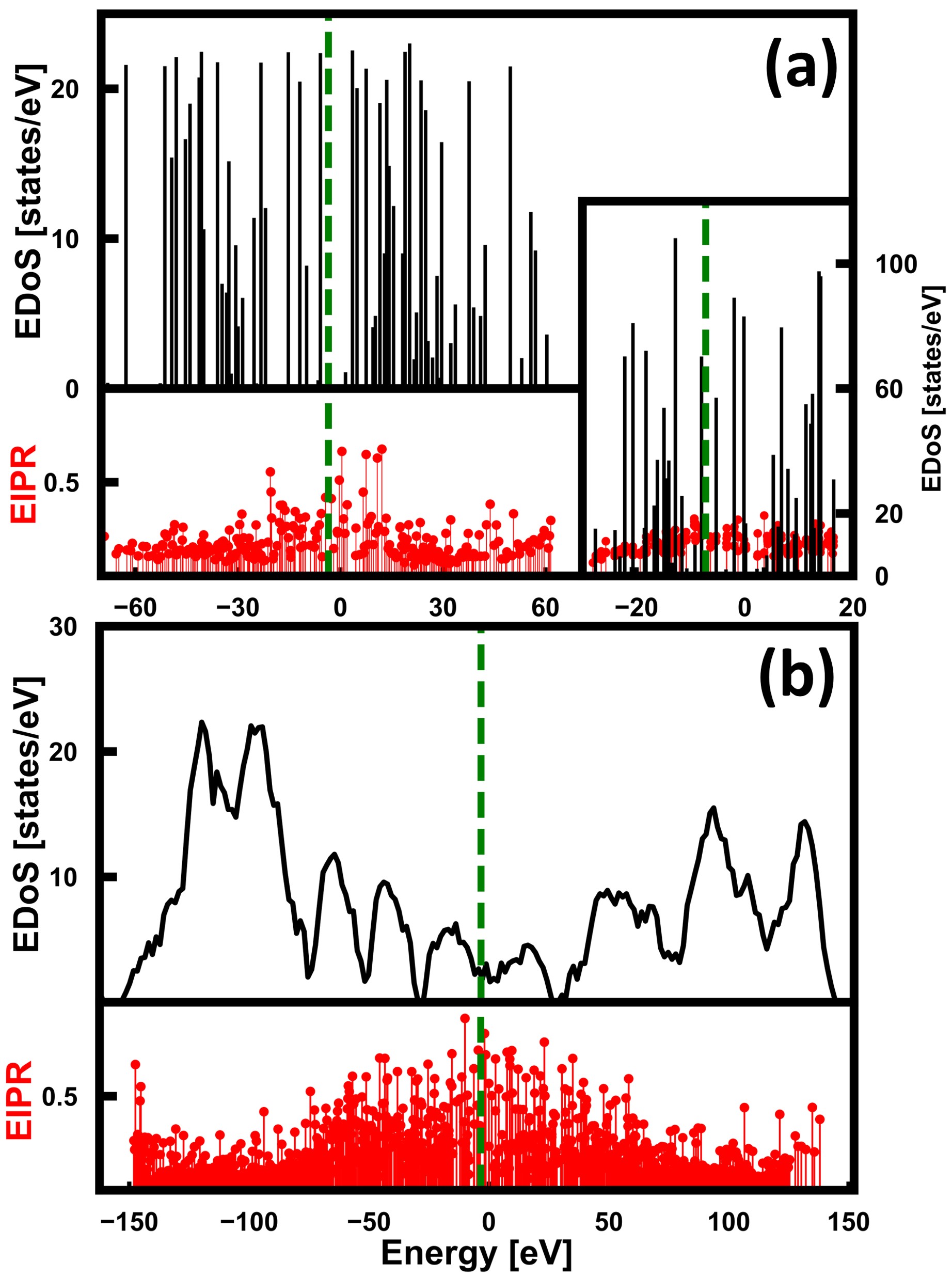}
	\includegraphics[width=\linewidth]{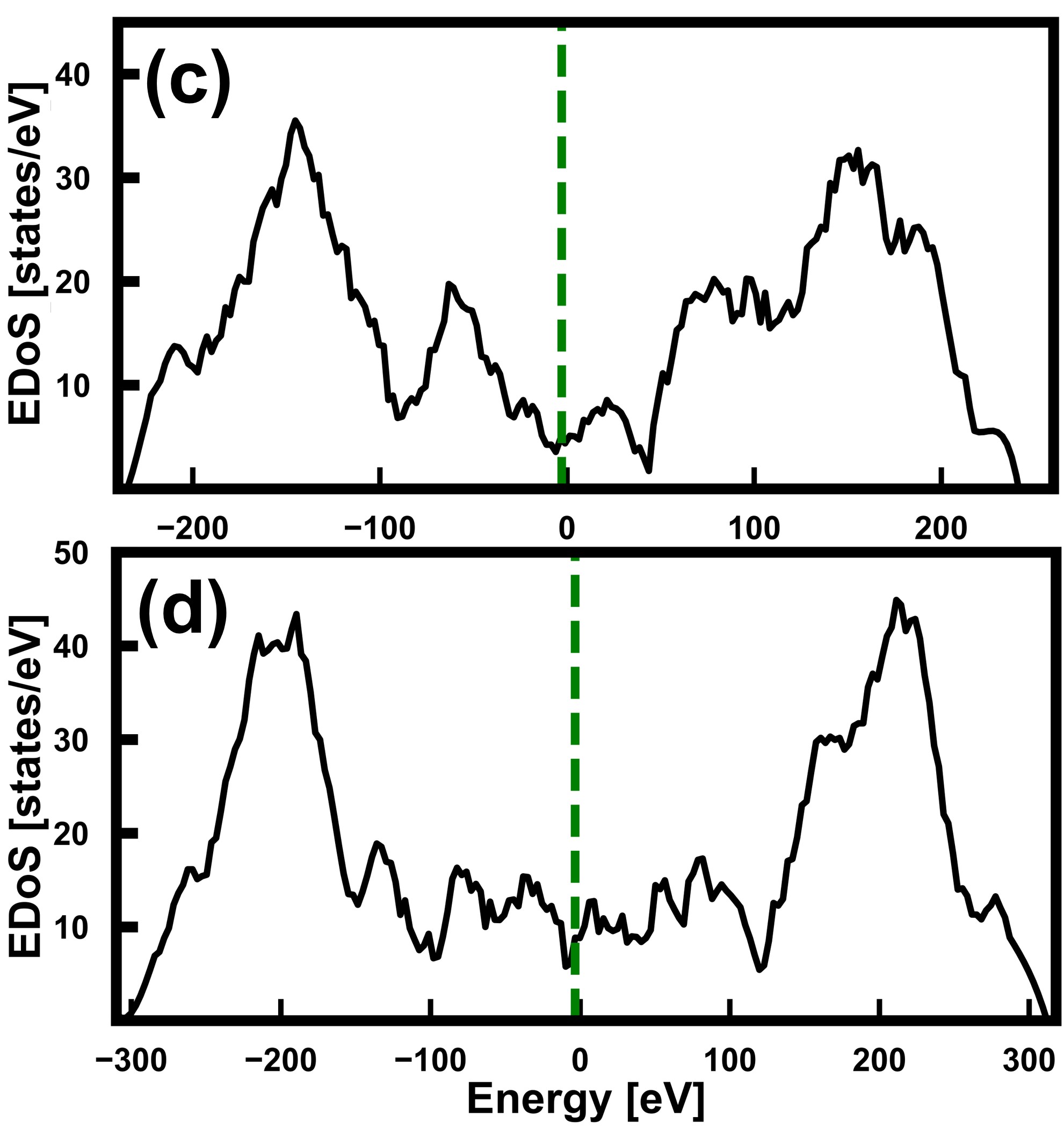}
	\caption{EDoS and EIPR for (a) BO$_{60}$, C$_{60}$ \cite{K_F60} (inset)  and  (b) BO$_{300}$.  EDoS was also computed for (c) BO$_{840}$ and  (d) BO$_{1374}$. The Fermi Energy for all plots is represented as a green vertical-dashed line. The calculations were implemented in \texttt{SIESTA}}
	\label{fig:Cfig_edos}
\end{figure}


 \begin{figure*}[!ht]
	\centering
	\includegraphics[width=\linewidth]{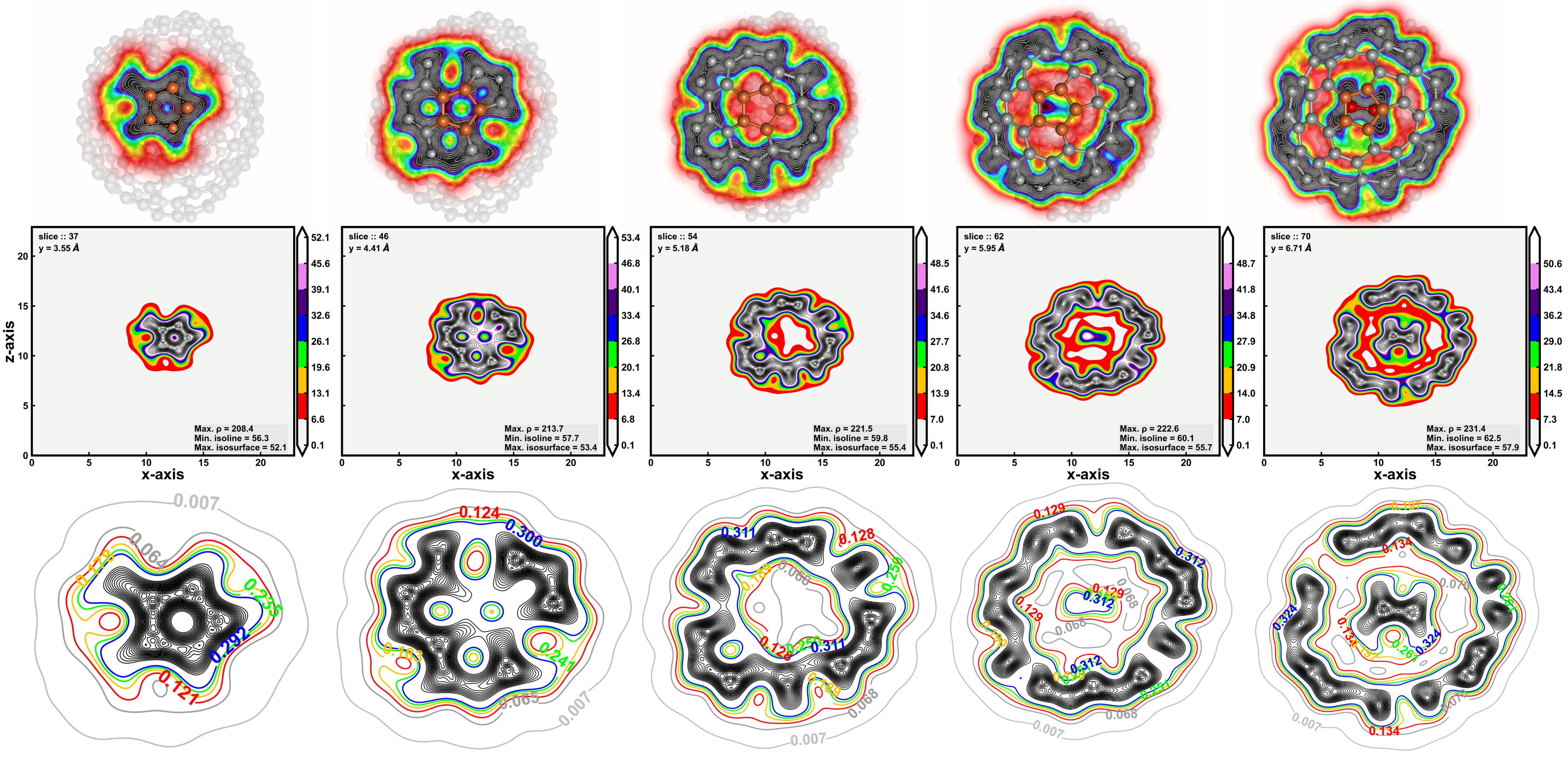}
	\caption{Cross-sections of the charge density [in $e^-$/\AA$^3$] distribution between 2 shells in a  BO$_{300}$ model. The charge density values were scaled by a factor of 100. The location of atoms, isosurface, and contour plots for 5 slices are presented in the first ($R_1$), second ($R_2$), and third ($R_3$) row. The columns ($C^1$ - $C^5$) show the progression from the outer to the inner shell. $C^3$ is the  mid-slice between $C^1$ and $C^5$, and $C^2$ ($C^4$) is the mid-slices between $C^3$ and $C^1$ ($C^5$). The charge distribution in the opposite end of the plane considered here is shown in the supplementary material \cite{suppl}.}
	\label{fig:Cfig_projectionSTART.jpg}
\end{figure*}

The electronic properties of the models were analyzed from calculations of the electronic density of states (EDoS) and the localization of Kohn-Sham states ($\phi$) was calculated as the corresponding electronic inverse participation ratio (EIPR), which is given by:

\begin{equation}
    I(\phi_n) =  \frac{\sum_i {| c_n^i |^4  }}{{(\sum_i {| c_n^i |^2} )^2}}
\end{equation}

\noindent where c$_n^i$ is the contribution to the eigenvector ( $\phi_n $) from the i$^{th}$ atomic orbital (s and p ) as implemented within \texttt{SIESTA} using the LDA self-consistent energy functional with a single $\zeta$ basis set. Low EIPR values correspond to highly extended states (evenly distributed over N atoms) and high EIPR values describe localized states. Fig. \ref{fig:Cfig_edos} (a) shows the EDoS and EIPR for the BO$_{60}$ model and pristine C$_{60}$ fullerene (inset in \ref{fig:Cfig_edos} (a)). While discrete states were observed for the C$_{60}$ and BO$_{60}$ models, the structural symmetry present in C$_{60}$ fosters degeneracy in its electronic density of states; therefore, electronic states occupying the same energy level (extended EIPR with low values) exists. On the other hand, the absence of certain structural symmetries in BO$_{60}$ resulted in moderately localized states over a wider energy range, breaking the electronic degeneracy. Similar high EIPR values were calculated for BO$_{300}$ (see Fig. \ref{fig:Cfig_edos} (b)), but unlike in BO$_{60}$, the states in BO$_{300}$ were distributed in a continuous spectrum suggesting non-degeneracy in its energy landscape.

The electronic density of states in the vicinity of the  Fermi energy ($E_f$) is important to many physical properties like transport and superconductivity. The conducting behavior of graphite is a consequence of the absence of any overlap in the valence (\text{$\pi$}) and conduction ($\pi^*$) bands at $E_f$. Drabold and co-workers, in their work on the spectral properties of large single-shell fullerenes \cite{DRABOLD1995833}, reported the narrowing of the $E_f$ gap as the number of atoms increases. In this work, the EDoS plot shows a gap for BO$_{60}$ at $E_f$, while for larger BO$_N$ models do not have any gap at $E_f$ (see Fig. \ref{fig:Cfig_edos} (c) and (d) for  BO$_{840}$ and BO$_{1374}$ respectively, with states engulfed around $E_f$). This behavior has been investigated by Liu and co-workers \cite{K_article}, using the scanning tunneling spectroscopy technique, from which they observed that small-$N$ buckyonions were semi-conducting, and increasing metallic properties were observed with increasing $N$, and hence with an increasing number of shells as well. 

Inter-layer cohesion in layered carbon structures like graphite and amorphous graphite derive from a combination of Van der Waals forces, as well as delocalized \text{$\pi$} electrons in the galleries \cite{36,LAG}. This non-dispersive contribution, involving weak ``metallic" interactions between the quasi-free electrons, was proposed by Rozp\text{$\l$}och and co-workers \cite{36,rozploch2003new}. Interpretation of the contribution of the \text{$\pi$} electrons to inter-layer cohesion requires a calculation of the individual contribution of the wavefunctions (Kohn-Sham orbitals) describing the system. Therefore, within the interpretation of the DFT formalism, the distribution of the delocalized electron gas in the gallery was investigated using electronic charge density distribution. The total charge density was calculated using both PBE and LDA functionals for BO$_{300}$ models. The obtained values were spatially projected on a  240$\times$240$\times$240 grid. The average charge density contribution from all the bands is presented in Table \ref{tab:Ctable_chgdensity}, and five cross-sections of the total charge distribution between two shells (in the -xz plane) are shown in Fig. \ref{fig:Cfig_projectionSTART.jpg}.

  \begin{table}[!ht]
	\begin{center}
		\caption{ Average PBE and LDA values for the total and \text{$\pi$} orbital charge densities for BO$_{300}$ on a 240-bin 3D grid.}\label{tab:Ctable_chgdensity}
		\begin{tabular*}{\linewidth}{@{\extracolsep\fill}ccccc@{\extracolsep\fill}}
			\hline
                \\
			 \textbf{Avg. Charge Density} &  \textbf{PBE [$e^-$/\AA$^3$]}  & \textbf{LDA [$e^-$/\AA$^3$]}\\
			\hline \\
			\textbf{All bands }  &     &   \\
   			Supercell :  & $2.199 \pm 0.035$   & $2.201 \pm 0.035$   \\
                Gallery  :  & $0.095 \pm 0.007 $   & $ 0.096 \pm 0.005$  \\\hline \\

                \textbf{\text{$\pi$} bands}  &     &   \\
   			Supercell :  & $0.543 \pm 0.037 $  & $0.548 \pm 0.037$   \\
                Gallery  :  & $0.021 \pm 0.001 $   & $0.020 \pm 0.001$  \\\hline  

		\end{tabular*}
	\end{center}
\end{table}

 \begin{figure}[]
	\centering
	\includegraphics[width=\linewidth]{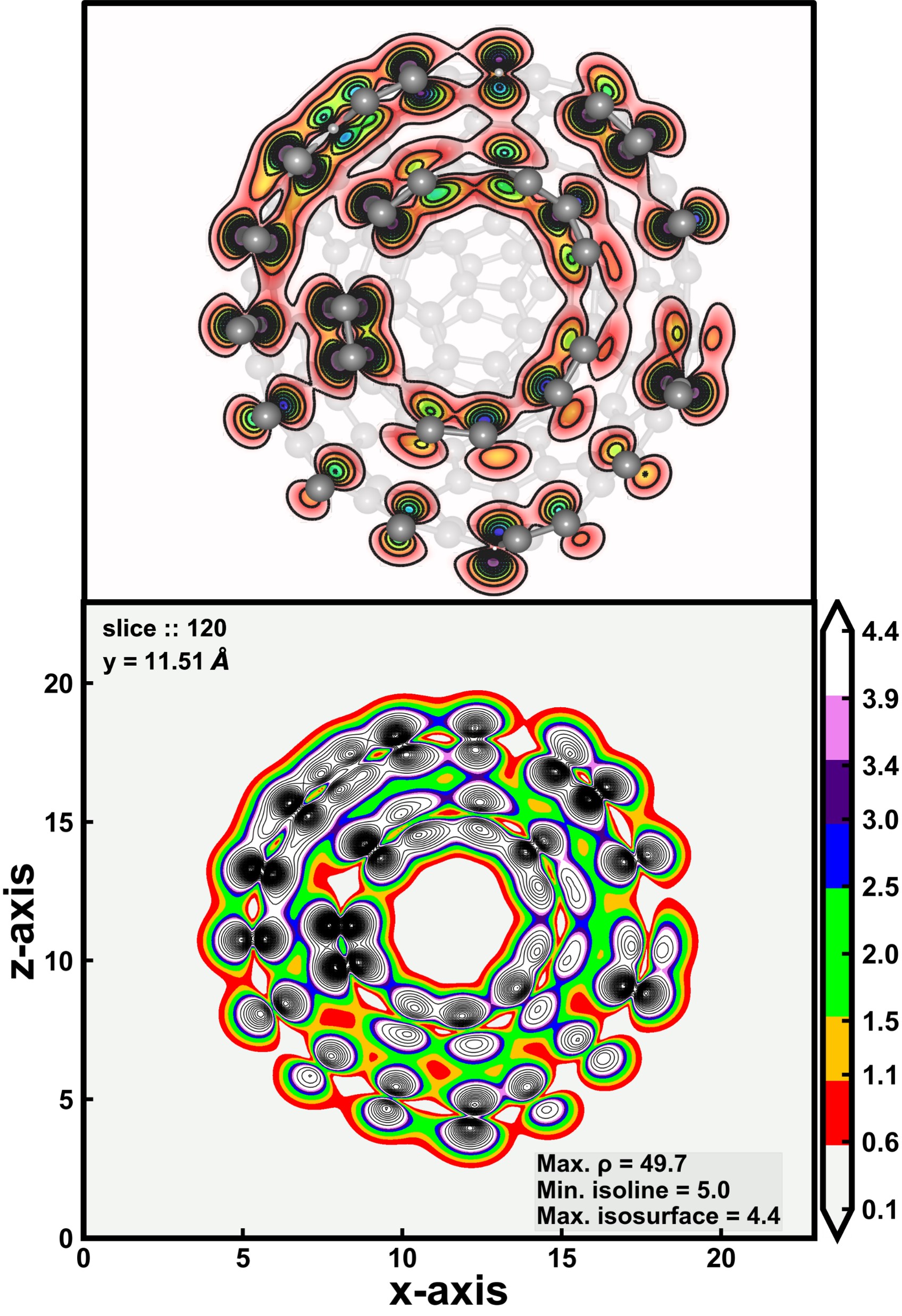}
	\caption{The delocalized \text{$\pi$}-electrons in the mid-section of the -xz plane in a BO$_{300}$ model. The PBE functional was used in this calculation and the charge density values were scaled by a factor of 100.}
	\label{fig:Cfig_PARCHG_PBE.jpg}
\end{figure}

In subsequent discussions on Fig. \ref{fig:Cfig_projectionSTART.jpg}, $R_jC^k$ represents the position defined by row $j$; column $k$. Using only $R_i$ ($C^i$) refers to all the columns (row) in the $i^{th}$ row (columns). $R_1$, $R_2$, and $R_3$ show the location of atoms, isosurface, and contour plots for each cross-sectional slice respectively. $C^1$ - $C^5$ show the progression from the outer to the inner shell. $C^3$ is the  mid-slice between $C^1$ and $C^5$, and $C^2$ ($C^4$) is the mid-slice between $C^3$ and $C^1$ ($C^5$). The charge density values in $R_2$ were scaled by a factor of 100. As a guide, the first set of atoms contributing to the charge density from both shells of the buckyonion, shown in $R_1C^1$ and $R_1C^5$ are colored in brown and red. Table \ref{tab:Ctable_chgdensity} shows that the average charge density in the mid-gallery region is $\approx$ $4.3 \%$ of the maximum charge density in the supercell. The low charge density values at the mid-gallery region are consistent with results from crystalline graphite (almost zero at the mid-gallery) and amorphous graphite (mid-gallery value was at least $2 \%$ of the maximum value of charge density in the system)\cite{LAG}. The higher values calculated for buckyonions could be a consequence of increased \text{$\pi$} electron mixing due to the local atomic curvature. For completeness, Fig. S4 shows the charge density plot for the last 5 cross-sectional slices at the opposite end of the model shown in Fig. \ref{fig:Cfig_projectionSTART.jpg}. Additionally, an animation (``TotalChgDensity.mp4") that shows the progression of the total charge density in the -xz plane along the y-direction is provided in the supplementary material. The animation confirms that the distribution of electrons in the gallery influences inter-shell cohesion in buckyonions and the charges are redistributed based on where atoms are located in each shell.

\begin{figure*}[!htbp]
	\centering
 \includegraphics[width=\linewidth]{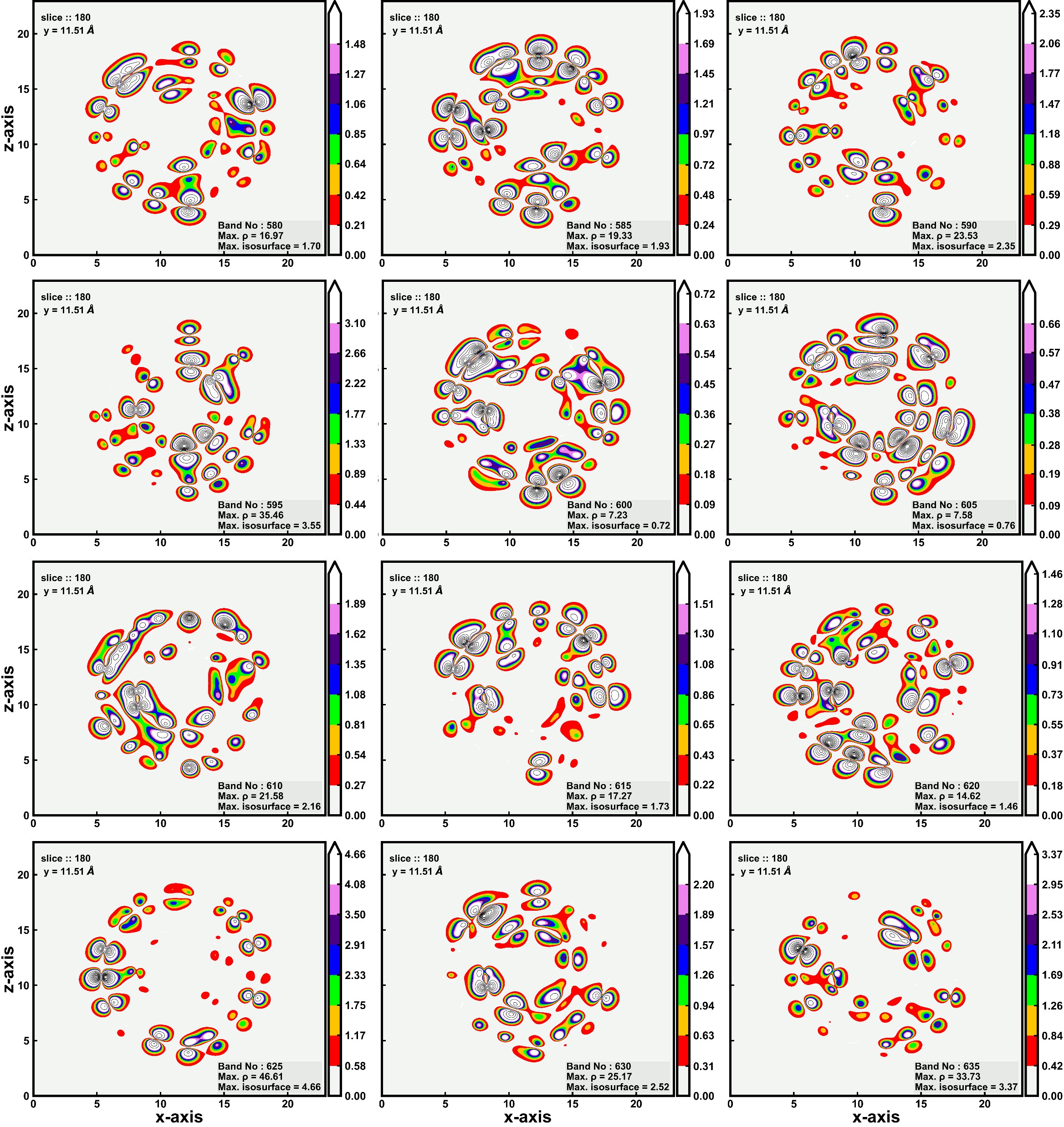}
	\caption{Single band decomposition of the $\pi$ electron bands close to the Fermi-level (in the 600$^{th}$ band). 4 (7) bands below (above) the Fermi band are plotted. The charge density values were scaled by a factor of 1000.}
 \label{fig:Cfig_orbital_mixing.jpg}
\end{figure*}

 Next, the actual electronic bands contributing to the inter-shell cohesion were obtained from the spatial band decomposition of the  $\pi$-orbitals from the total charge density in the same 3D 240-grid. The $\pi$ electron charge density at the mid-section of the -xz plane for a BO$_{300}$ model is shown in Fig. \ref{fig:Cfig_PARCHG_PBE.jpg}. The charge density values were scaled by a factor of 100. The locations of the contributing atoms are shown in Fig. \ref{fig:Cfig_PARCHG_PBE.jpg} [TOP]. The values for the $\pi$ electron charge density in the gallery $\approx$ 0.023 $e^-$/\AA$^3$, and increases to $\approx$ 0.033 $e^-$/\AA$^3$ for atoms in each shell that are positioned opposite each other. In general, the average $\pi$ electron charge density in the gallery is $\approx$ 4 \%  of the average in the supercell (see Table \ref{tab:Ctable_chgdensity}). An animation showing the distribution of the $\pi$ electrons in a BO$_{300}$ model is also provided in the supplementary material. For comparison, the total and $\pi$ band charged densities, calculated using LDA functional, showed close values to the results obtained using PBE functional (see Table \ref{tab:Ctable_chgdensity}). This is also reflected in the close similarity between the isosurface plot in Fig \ref{fig:Cfig_PARCHG_PBE.jpg} and its LDA-calculated counterpart in Fig S5 \cite{suppl}. Finally, the contribution of individual $\pi$ bands to inter-layer cohesion was also investigated. The plots in Fig. \ref{fig:Cfig_orbital_mixing.jpg}, which has charge density values scaled by a factor of 1000, show the isosurface of 4 (7) bands below (above) the 600$^{th}$ band which contains the Fermi level (Fermi band), which corresponds to the $\pi$ ($\pi$*) orbitals. To obtain finer isosurfaces, the grid bin size was increased to 360$\times$360$\times$360.  The average contribution of the $\pi$* electrons in the gallery is (at most) 2 orders of magnitude less than values from the $\pi$ bands which means that the delocalized $\pi$ electrons are  the  major contributions to the inter-layer cohesion. The results also show that contributions from more than one band can be localized in a particular atom which means that the $\pi$ bands are linear combinations (mixing) of the  wavefunctions (Kohn-Sham orbitals) that describe the buckyonions. It is noteworthy that in these non-classical fullerenes, layering persists despite the absence of an exact (graphitic) stacking registry. 

\section{Conclusion}
In this work, the formation of buckyonions from initial random configurations of carbon atoms was achieved using the GAP potential. The structure formed within 100 ps and DFT energy validation revealed an energy difference in the range of 0.02 - 0.08 eV/atom using \texttt{VASP} and \texttt{SIESTA}. Multi-shell fullerenes  with  1 $\sim$ 4 layers were created for atoms ranging from 60 $\sim$ 3774 atoms. While there remains a level of topological disorder in each shell (i.e Stone-Waals defects), the number of atoms in the innermost shell remained close to 60 or 180 atoms. The buckyonions prefer to form starting with the  outermost shell and then build inwards with a consistency in gallery width of $\approx$ 2.87 \AA~between the two outermost layers. Clustering was proposed as a formation mechanism for the buckyonions. The electronic density of states calculation showed that more states appear in the Fermi Level as $N$ increases. Finally, the physics of inter-shell layering was investigated by considering the charge density in the \text{$\pi$} subspace. Results showed that delocalization of the \text{$\pi$} electron led to an electronic charge cloud in the gallery, which causes cohesion between shells. 

\begin{acknowledgments}
The authors thank Anna-Theresa Kirchtag for proofreading the manuscript, the U.S. Department of Energy for support under Grant No. DE-FE0031981, and XSEDE (supported by National Science Foundation Grant No. ACI-1548562) for computational support under allocation no. DMR-190008P
\end{acknowledgments}

\newpage

\bibliography{BO.bib}
\bibliographystyle{unsrt}

\end{document}



\title{Supplemental material: Simulation of multi-shell fullerenes using Machine-Learning Gaussian Approximation Potential}

\author{C. Ugwumadu}
\email{cu884120@ohio.edu}
\author{K. Nepal}
\author{R. Thapa}
\author{Y. G. Lee}%

\affiliation{Department of Physics and Astronomy, \\
Nanoscale and Quantum Phenomena Institute (NQPI),\\
Ohio University, Athens, Ohio 45701, USA}%

\author{Y. Al Majali}%
\author{J. Trembly}%
\affiliation{Russ College of Engineering and Technology, \\
Ohio University, Athens, Ohio 45701, USA}%

\author{D. A. Drabold}%
\email{drabold@ohio.edu}
\affiliation{Department of Physics and Astronomy, \\
Nanoscale and Quantum Phenomena Institute (NQPI),\\
Ohio University, Athens, Ohio 45701, USA}%

\date{\today}

\maketitle

\section{Description of Animations produced for the buckyonion models}
To aid in visualizing some of the discussions in the paper, we have produced some animations for some buckyonion models. The animations can also be found \href{https://people.ohio.edu/drabold/cepa_movies/}{here} or by visiting the url: \url{https://people.ohio.edu/drabold/cepa_movies/}.

\begin{enumerate}
  \item BO300\_mov.mp4 and BO1374\_mov.mp4 files show the buckyonion formation process for  BO$_{300}$ and  BO$_{1374}$ respectively. The outer, middle and inner shells are colored green, red, and blue in the BO1374\_mov.mp4 file respectively. The  green (red) colored atoms in the BO300\_mov.mp4 file represent the outer (inner) shell. 
  
  \item BO\_growthProcess.mp4 shows the clustering and growth process for 540 randomly distributed atoms within a 720-atom fullerene isomer. The shells are colored the same as described for the BO1374\_mov.mp4 file. The pentagon rings in the 720-atom fullerene isomer are colored black to show their positions.

  \item TotalChgDensity.mp4 and PiChgDensity.mp4 shows the progression of the charge density in the -xz plane for all bands and only the $\pi$-bands respectively. T The mid-section discussed in the manuscript is at y = 11.51 \AA 
\end{enumerate}

\section{Supporting Tables and Figures}
 
\begin{figure}[!ht]
	\includegraphics[width=.9\linewidth]{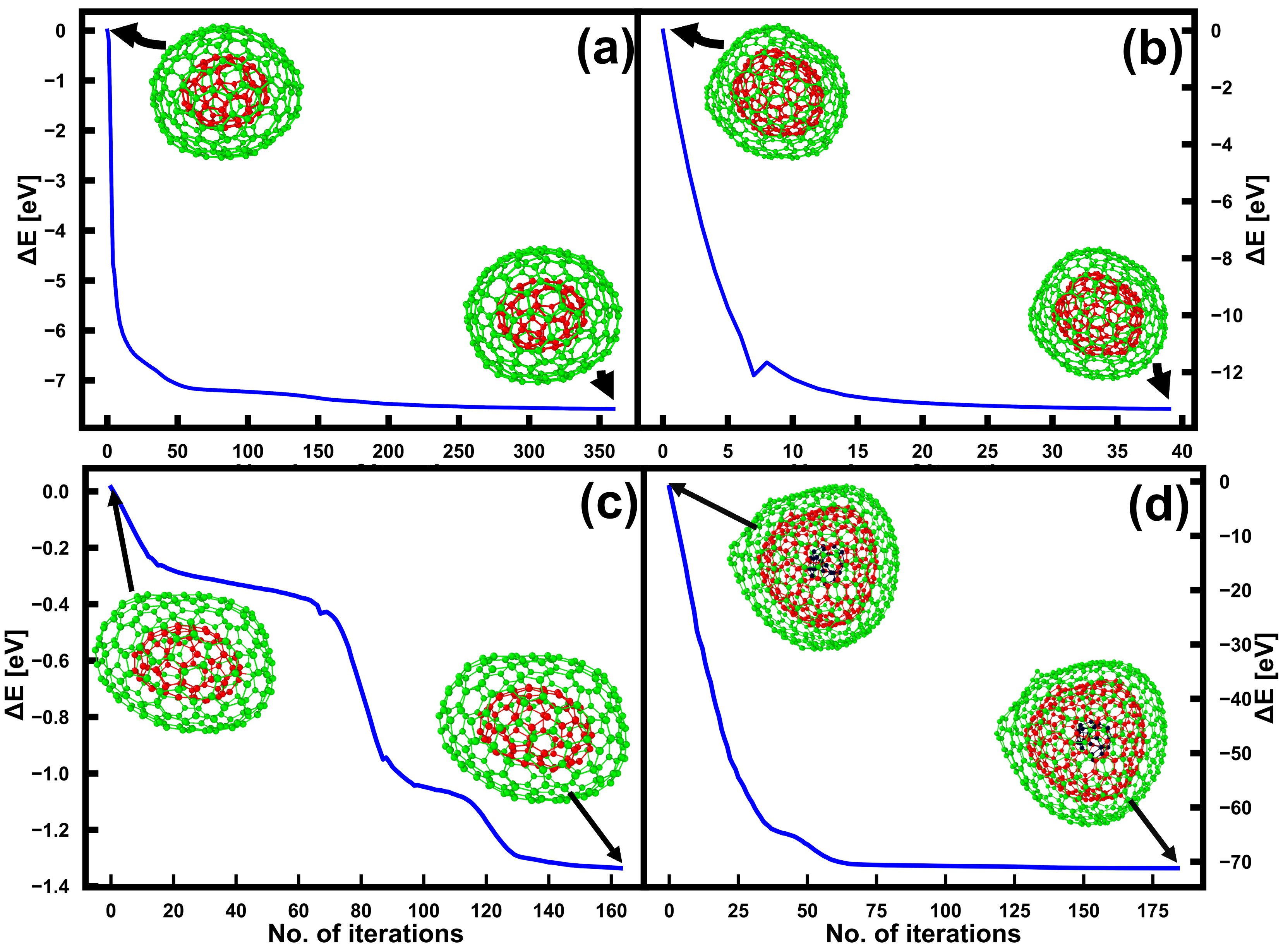}
	\caption{Plots of the energy difference ($\Delta$E) against the Conjugate gradient (CG) iterations carried out using VASP on BO$_{300}$  (a) and BO$_{540}$ (b). SIESTA was used for the BO$_{300}$ (c) and BO$_{840}$ (d) models as well. The insets in a-d represent the models before and after CG relaxation.}
	\label{fig:CSfig_VASP_SEISTA_CG}
\end{figure}

\begin{figure}[!ht]
	\centering
	\includegraphics[width=.9\linewidth]{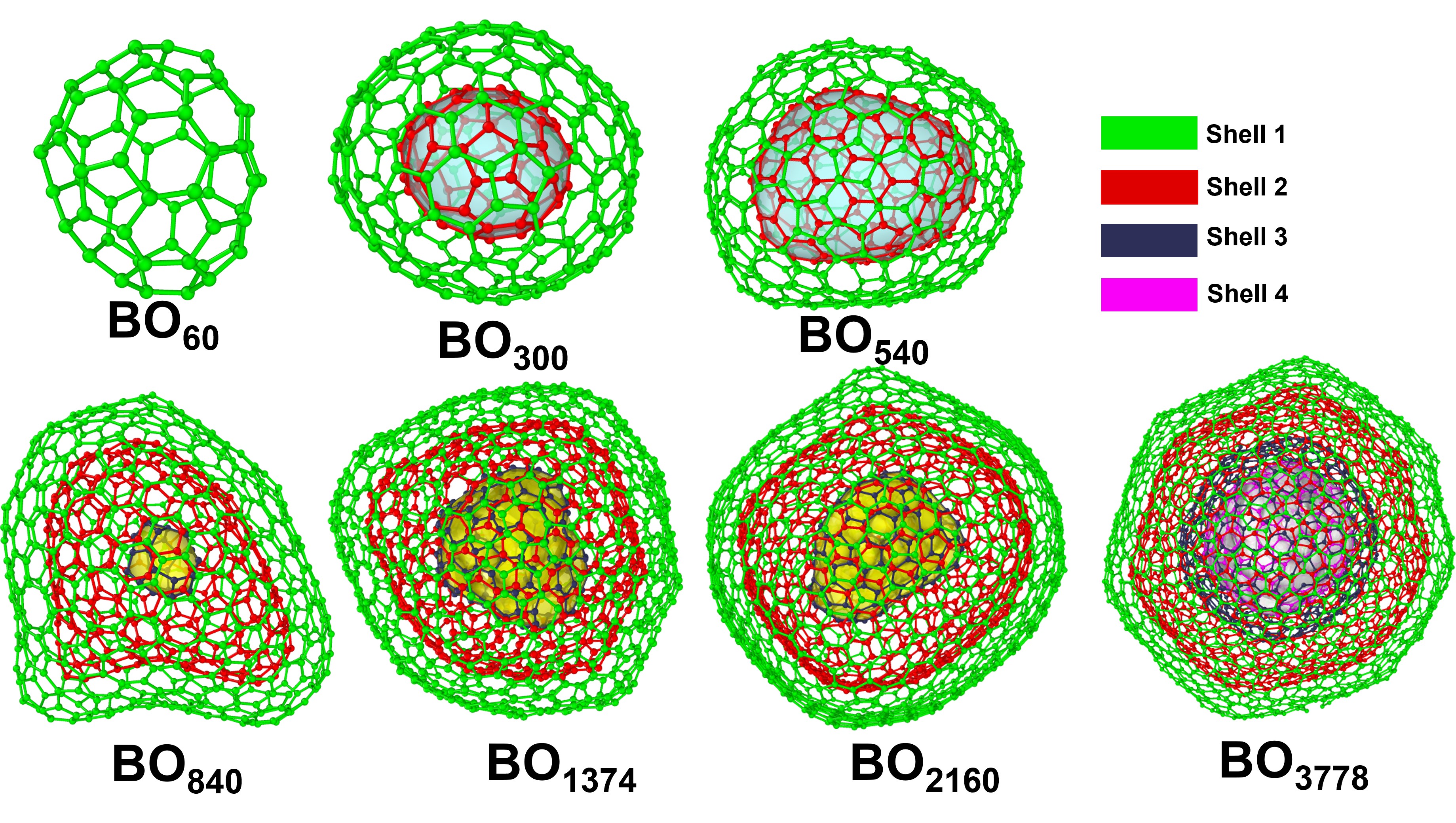}
	\caption{Some of the buckyonions models simulated in this work}
	\label{fig:figS1}
\end{figure}

\begin{table}[!htbp]
	\caption{Ring-size analysis for the buckyonion models. }
		\label{tab:Ltab_ring_number}
	\begin{ruledtabular}
		\begin{tabular*}{0.6\linewidth}{@{\extracolsep\fill}c|c|ccccc}
			\textbf{Models}&   \textbf{Layers}& & & \textbf{Ring size:} & & \\
			\hline
			\hline
			  & & \textbf{5} & \textbf{6} & \textbf{7} & \textbf{8} & \textbf{9}\\
			\hline
			\textbf{BO$_{60}$} &$s_1$&  12 & 20 & 0 & 0 &0  \\
			\hline

			{\textbf{BO$_{300}$}} & $s_1$  & 15 & 22 & 3 & 0 & 0 \\
			 &  $s_2$   & 26 & 74 & 14 & 0 & 0 \\
			\hline
			
			{\textbf{BO$_{540}$}} &  $s_1$    & 26 & 50 & 12 & 1 & 0 \\
			 &$s_2$   & 62 & 85 & 28 & 8 & 2  \\
			\hline

			{\textbf{BO$_{840}$}} &   $s_1$ & 6 & 1 & 0 & 2 & 1\\
			&  $s_2$  & 34 & 84 & 20 & 3 & 0\\
			&  $s_3$ & 68 & 148 & 40 & 11 & 0\\
			\hline
			
			{\textbf{BO$_{1374}$}} &  $s_1$  & 27 & 34 & 14 & 3 & 1\\
			& $s_2$ & 25 & 189 & 13 & 2 & 0\\
			& $s_3$  & 50 & 292 & 36 & 3 & 0 \\
			\hline
			
			{\textbf{BO$_{2160}$}} &  $s_1$  & 15 & 20 & 10 & 2 & 1\\
			& $s_2$ & 36 & 291 & 25 &0 & 5\\
			& $s_3$  & 55 & 380 & 32 & 2 & 5 \\
			\hline

		    {\textbf{BO$_{3774}$}} &  $s_1$  & 31 & 91 & 17 & 4 & 4 \\
		    &  $s_2$ & 47 & 183 & 26 & 5 & 4 \\
			&  $s_3$ & 112 & 333 & 80 & 18 & 6\\
			& $s_4$& 179 & 422 & 123 & 24 & 7\\
		\end{tabular*}
	\end{ruledtabular}
\end{table}

\begin{figure}[!ht]
	\includegraphics[width=.9\linewidth]{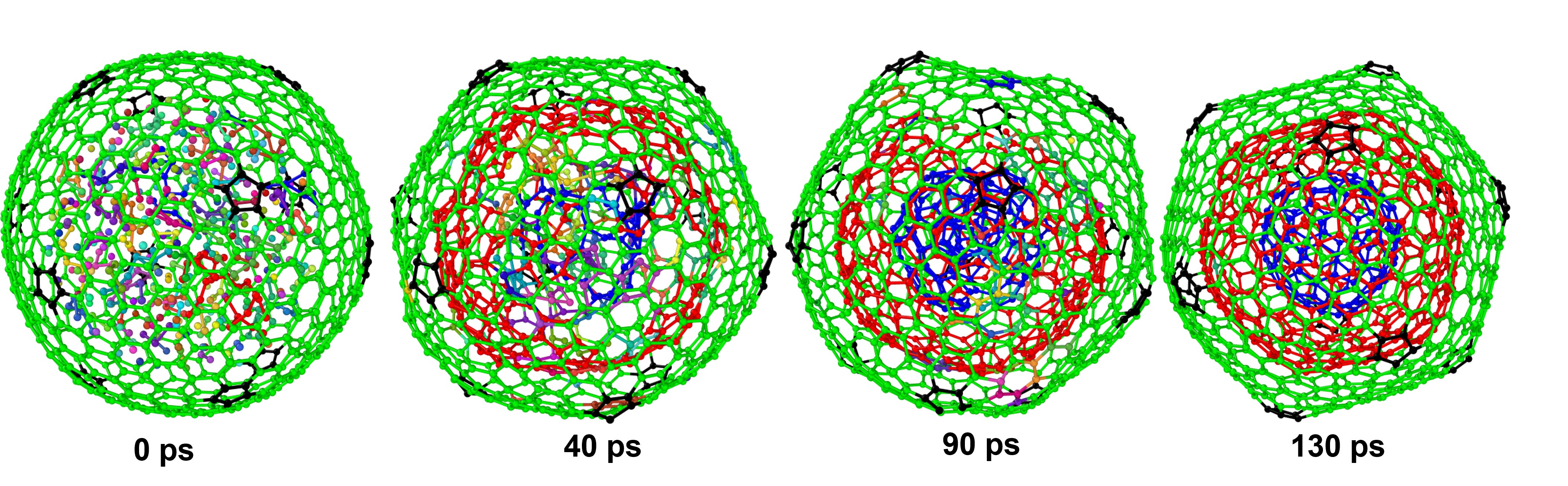}
	\caption{Growth Process from C$_{720}$ isomer with 540 atoms randomly distributed C atoms. The outermost shell (green) remained with 720 atoms at the end of the simulation. The heptagons in the outermost shell are colored black.}
	\label{fig:CSfig_740_520_formation}
\end{figure}

\begin{figure}[!ht]
	\centering
	\includegraphics[width=\textwidth]{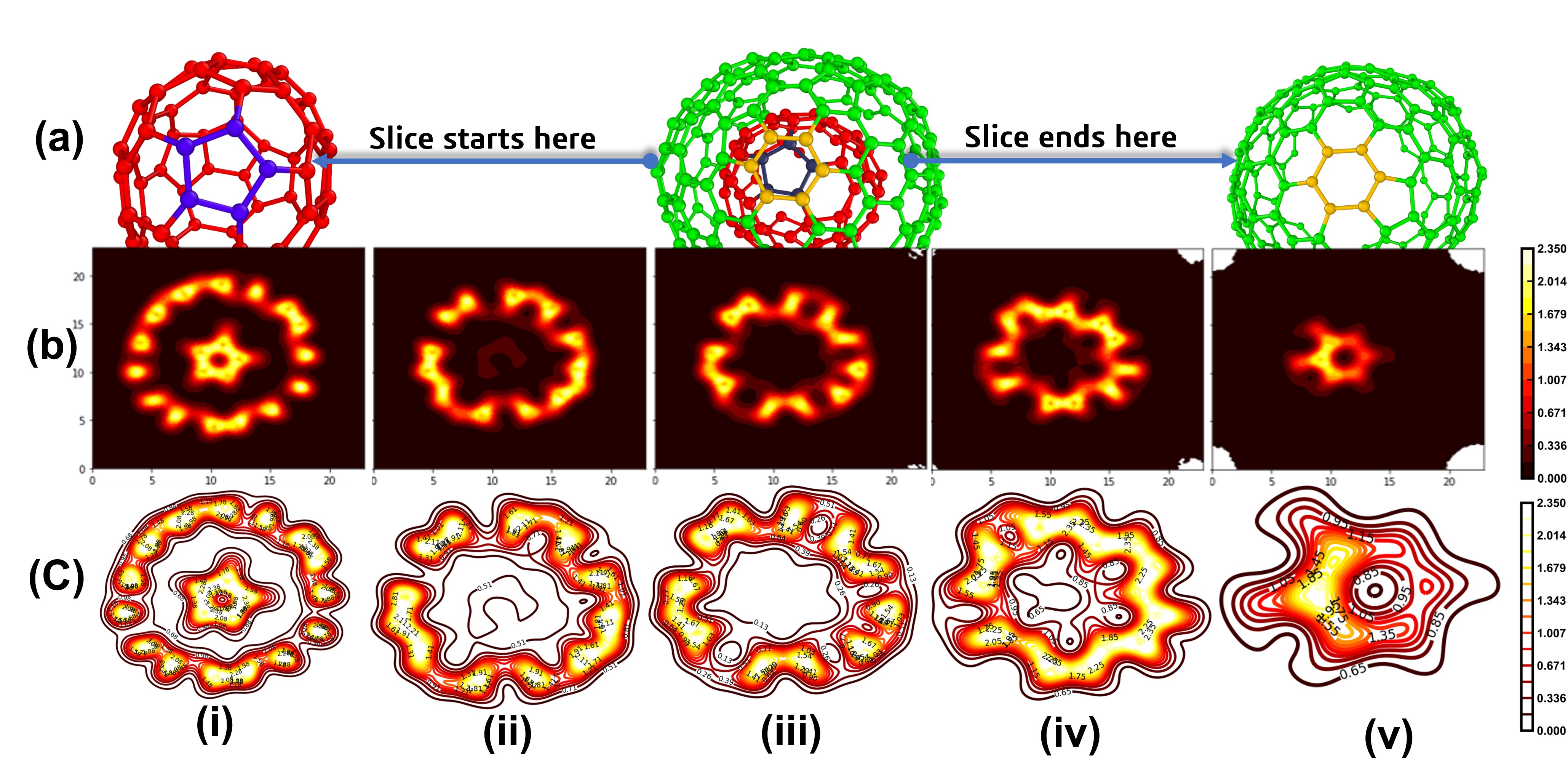}
	\caption{The distribution of the charge density [in $e^-$/\AA$^3$] for other end of the BO$_{300}$ model considered in Fig. 8 in the main manuscript. In the first row (a), the first column (a (i)) starts with a pentagon (yellow) in the outer shell (green), and the fifth column (a(v)) first slice in the inner shell (red) showing 2 carbon atom. The second and third rows (b and c) show the heat map (b (i - v) and contour values (c (i - v)) of the charge density for 5 slices in columns i - v. The charge distribution in the opposite end of the plane considered here is discussed in the main manuscript.}
	\label{fig:figS2}
\end{figure}

\begin{figure}[!ht]
	\centering
	\includegraphics[width=\textwidth]{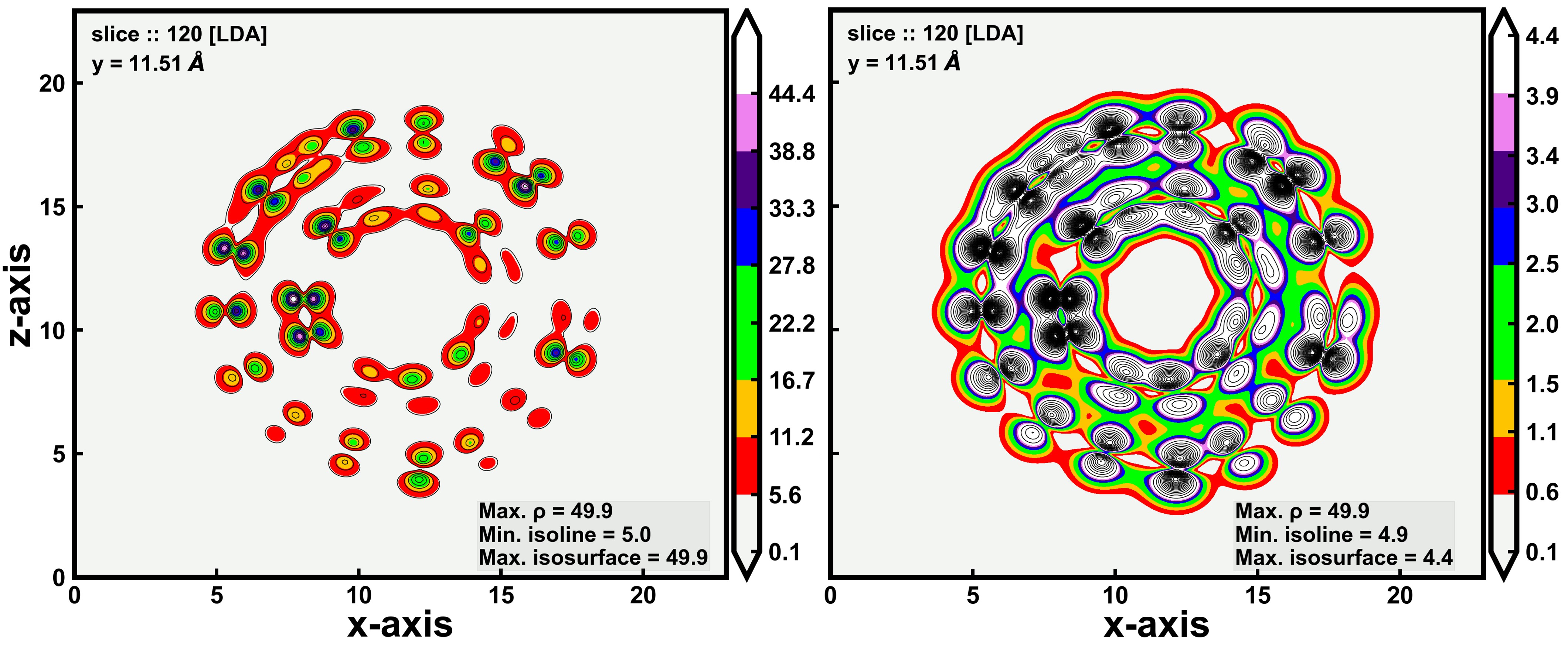}
	\caption{The delocalized \text{$\pi$}-electrons in the mid-section of the -xz plane in a BO$_{300}$ model using LDA functional.The values of the charge density were scaled by a factor of 100. The left plot shows the entire range of $\pi$ charge density values in that particular slice and the atoms are positioned in the center of the dumbbell shape which is formed with contour lines with a minimum threshold cutoff of 0.034 $e^-$/\AA$^3$. The light-blue region in-between the shells (gallery) suggests that the value of the delocalized electron charge density is less than 0.074 $e^-$/\AA$^3$. The left plot indicates that the value $\pi$ electron charge density in the gallery ranges majorly between 0.012 to 0.023 $e^-$/\AA$^3$ and has values up to 0.034 $e^-$/\AA$^3$ when there are atoms that are positioned opposite each other in different shells.}
	\label{fig:CSfig_PARCHG_LDA.jpg}
\end{figure}

